# THE PUZZLE OF COMPOSITION OF COSMIC RAYS WITH ENERGIES (2 - 12.5) EEV ACCORDING TO MUON DETECTORS DATA OF THE YAKUTSK EAS ARRAY

**A.V. Glushkov[1], L.T. Ksenofontov[1,*], K.G. Lebedev[1], A.V. Saburov**

[1] *Yu.G. Shafer Institute of Cosmophysical Research and Aeronomy SB RAS, Yakutsk, Russia*

The results of a study of the cosmic ray composition in individual events in the energy range (2–12.5) EeV using the muon correlation method [31] is presented. The considered sample included showers with zenith angles less than 60 degrees recorded in the period 1974–2018. The existence of four separate groups of primary particles with different origins is confirmed. The obtained results have potential importance for understanding the composition of cosmic rays in the specified primary energy range.

## 1. INTRODUCTION

The composition of ultra-high-energy cosmic rays (CR) ($E \geq 10^{15}$ eV) can only be studied using extensive air showers (EAS). It has been actively studied worldwide for over 50 years [1], but is still not precisely known. Without knowledge of the mass and charge composition of CR, it is difficult to understand the nature of nuclear interactions in this energy range and the sources of the primary particles. Various EAS parameters are used to study them. At the Yakutsk array, this is done using the lateral distribution functions (LDF) of the electron, muon, and Cherenkov components of the showers (see, for example, [2-7]). The key to solving the problem of CR composition is a simple relationship that follows from the principle of nucleon superposition [8]:

$$\ln A = \frac{d_{\mathrm{p}} - d_{\mathrm{exp}}}{d_{\mathrm{p}} - d_{\mathrm{Fe}}} \times \ln 56\,, \qquad (1)$$

where $A$ is the atomic number of the primary particle, $d$ is any parameter sensitive to the CR composition, obtained experimentally (exp) and by calculation for primary protons (p) and iron nuclei (Fe). In [9], the LDFs of the responses of the surface ground-based scintillation detectors (SD) and muon detectors (MD) of the Yakutsk array from primary particles with an energy of $E \geq 10^{17}$ eV were calculated using the QGSjet-01-d [10], QGSjet–II-04 [11], EPOS-LHC [12], and SIBYLL-2.1 [13] models within the framework of the CORSIKA software package [14].

Figure 1 shows the estimates of the primary particle composition by the world's EAS arrays. They show a contradictory picture resulting from the analysis of the average EAS characteristics.

*) e-mail: ksenofon@ikfia.ysn.ru

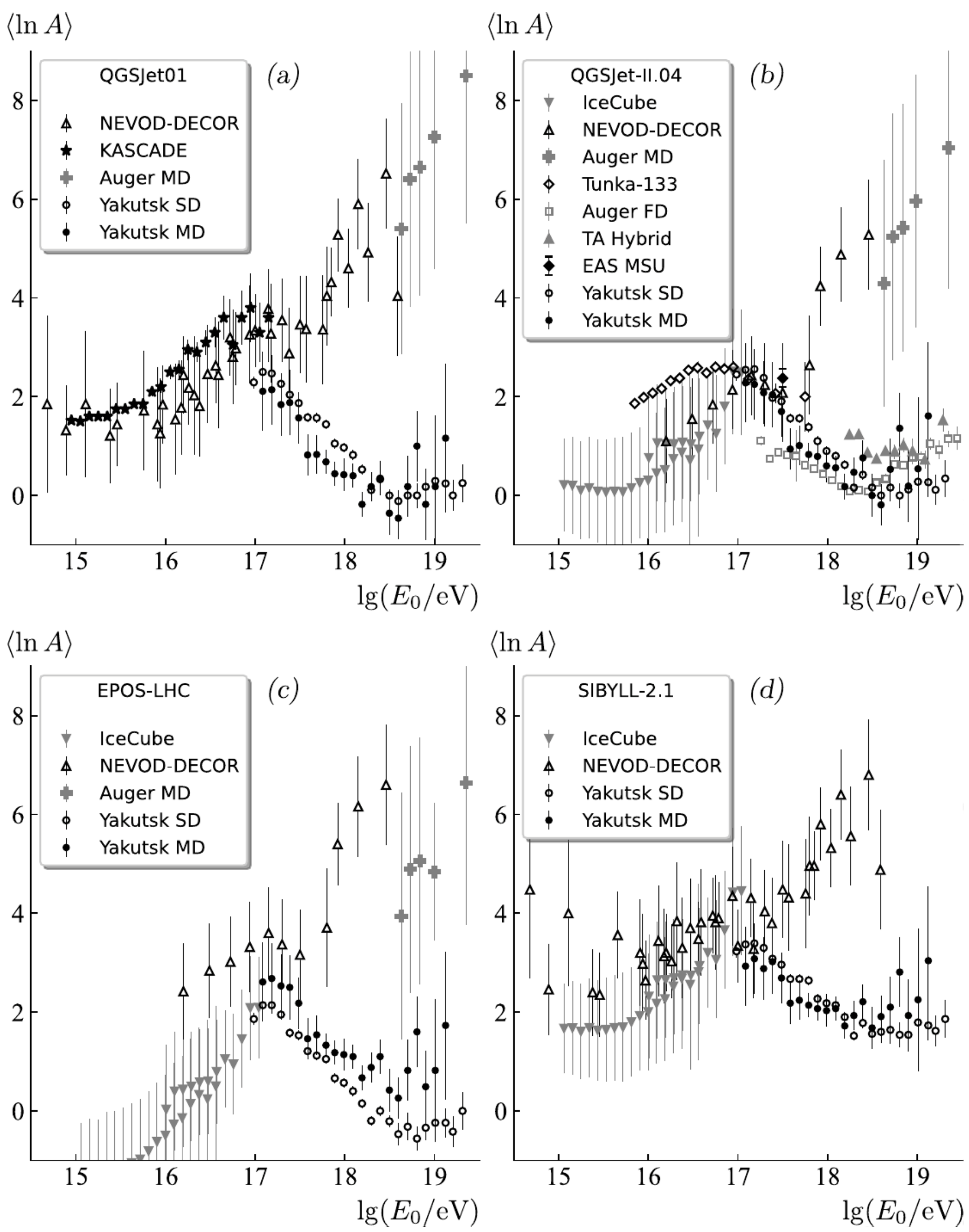

**Fig. 1.** Energy dependence of the CRs mass composition by the different EAS experiments. Open circles are Yakutsk surface detectors data (SD) [15, 16], solid circles are estimations by the muon component of EAS (μ) [17]. The estimates derived from the scaling parameter z [18] for the IceCube [19], NEVOD-DECOR [20, 21], PAO (Auger) [23–25] and EAS-MSU [22] experiments, the data by KASCADE [26], Tunka-133 [27] experiments and the fluorescent part of PAO (FD) [28] and TA [29, 30] are also presented here.

The Yakutsk array has accumulated significant experimental material, which is periodically analyzed as ideas about the nature of CR change and new methods for studying them appear. In this regard, we analyze the MD data of the Yakutsk array, which have large statistics and good measurement accuracy. A new muon correlation method for determining the mass composition of CR in individual events was proposed in [31]. It was tested on a sample of 127 EAS with an energy of $E \geq 1.25\times10^{19}$ eV. This method studies the correlation parameter:

$$\eta = \frac{S_{\mu}^{\exp}(E,\theta,r)}{S_{\mu}^{\text{sim}}(p,E,\theta,r)}, \tag{2}$$

where $S_{\mu}^{\exp}(E,\theta,r)$ is the density of MD responses in showers with energy $E$ and zenith angle $\theta$ at a distance from the axis $r$, measured in the experiment; $S_{\mu}^{\text{sim}}(p,E,\theta,r)$ is the same value obtained as a result of simulating EAS using one or another model of hadron interactions. The distribution of the values of this parameter in the sample of events [31] has four peaks. Two of them can be conditionally identified with primary protons and iron nuclei. The third is formed by several events with an abnormally high muon content. And, finally, the fourth peak is formed by EAS with an abnormally low muon content. Below are the results of studying the CR composition using the method [31] at energies $E \approx (2–4)\times10^{18}$eV.

## 2. EXPERIMENT AND DATA ANALYSIS METHOD

### 2.1. Characteristics of the array

The Yakutsk EAS array has been conducting continuous observations since January 1974. Shower events are recorded by SDs with an area of $2\times2$ m$^2$. The muon component of EAS is measured by underground MDs with a threshold energy above $1.0\times\sec\theta$ GeV. The first two MDs with an area of 36 m$^2$ began operating in the fall of 1978. In 1986, three more MDs with an area of $2\times10 = 20$ m$^2$ were put into operation. The arrival direction of EAS is reconstructed from relative delays of SD triggering within the approximation of flat shower front. Axis location is done with the use of lateral distribution function (LDF) in the form:

$$S(\theta,r) = S_{600}^{\exp}(\theta)\frac{600}{r}\left(\frac{600 + r_M}{r + r_M}\right)^{b_s(\theta)-1}, \tag{3}$$

which was derived at the Yakutsk array in the first years of its operation [32]. The parameter $S_{600}^{\exp}(\theta)$ characterizes the density of the SD responses at axis distance $r$ = 600 m obtained in experiment; $r_M$ is the Moliere radius [32, 33], which depends on the temperature $T$(K) and the atmospheric pressure $P$(mb):

$$r_M \approx \frac{7.5\times10^4}{P}\times\frac{T}{273}. \tag{4}$$

The value of $r_M$ for each shower is determined individually. Its average annual value for Yakutsk is $\langle r_M\rangle \approx 70$ m; $b_s$ is a structural parameter [32]:

$$b_s(\theta) = 1.38 + 2.16\times\cos\theta + 0.15\times\log10\left(S_{600}^{\exp}(\theta)\right). \tag{5}$$

The arrival direction, coordinates of shower axis and $S_{600}^{\exp}(\theta)$ is determined by minimizing the corresponding $\chi^2$. The primary energy $E$ is estimated by the relations [34]:

$$E = E_1\times S_{600}(0°)^B \text{ [eV]}, \tag{6}$$

$$S_{600}(0^\circ) = S_{600}^{\exp}(\theta)\times\exp\left(\frac{(\sec\theta{-}1)\times1020}{\lambda}\right)\ [\mathrm{m}^{-2}], \tag{7}$$

$$\lambda\ =\ (450\ \pm 44) + (32 \pm 15)\times\log_{10}\big(S_{600}(\theta)\big)\ \left[\frac{\mathrm{g}}{\mathrm{cm}^2}\right], \tag{8}$$

where $E_1 = (3.76 \pm 0.3)\times10^{17}$ eV and $B = 1.02 \pm 0.02$.

### 2.2. Lateral distribution of particles in EAS.

The calculation procedure of the used below SD and MD responses LDFs is described in [9]. The application of this technique indicates a certain agreement between the QGSjet01 [10] and QGSjet-II.04 [11] models of ultra-high energy hadron interactions and the experiment (see, for example, [15–17, 34–38]). As a result, a method for estimating the CR mass composition in individual showers has been developed [31]. It is based on a comparative analysis of the experimental LDFs with the values obtained during the EAS simulation using the CORSIKA package [14] within the framework of the QGSjet-II.04 [11] model for primary protons. The FLUKA2011 generator [39] was used in these calculations to simulate hadron interactions with energies below 80 GeV. The approximation of the SD LDF within the axis distance range оси $r \approx$ 30–2000 m has an analytical form [9]:

$$S(E,\theta,r) = S_{600}^{\exp}(E,\theta)\times\left(\frac{610}{r+10}\right)^{2}\times\left(\frac{608}{r+\ 8}\right)^{b_s(\theta,E)-2}\times\left(\frac{600+r_1}{r+\ r_1}\right)^{11}, \tag{9}$$

where $r_1 = 7000$ m. The parameter $b_s(\theta,E)$ depends on the primary energy:

$$b_s(\theta,E) = \beta_S(\theta) + 0.118\times(\log_{10}E{-}19) \tag{10}$$

and the zenith angle:

$$\beta_s(\theta) =\ a_s - f_s\times(\sec\theta{-}\ c_s). \tag{11}$$

The parameters in the eq. (11) for $\sec\theta \leq 1.25$ are $a_s = 2.65$, $f_s = 1.3$ and $c_s = 1.25$; for $1.25 < \sec\theta \leq 1.7$: $a_s = 2.65$, $f_s = 1.84$ and $c_s = 1.25$; and for $\sec\theta > 1.70$: $a_s = 1.83$, $f_s = 1.3$ and $c_s = 1.70$.

The LDF of muons in the distance range $r \approx$ 30–2000 m is well described by function [9]:

$$S_\mu^{\mathrm{sim}}(E,\theta,r) = S_\mu(E,600)\times\left(\frac{600}{r}\right)^{0.75}\times\left(\frac{600+r_0}{r+r_0}\right)^{b_\mu(\theta)-0.75}\times\left(\frac{600+r_1}{r+\ r_1}\right)^{9} \tag{12}$$

with $r_0 = 280$ m and $r_1 = 2000$ m. The parameter $b_\mu(\theta)$ depends only on the zenith angle:

$$b_\mu(\theta) = 1.15 - \beta_\mu \times (\sec\theta - 1.25). \tag{13}$$

If $\sec\theta \leq 1.25$ the parameter $\beta_\mu = 1.25$, otherwise $\beta_\mu = 1.6$. The parameter $S_\mu(E,600)$ in eq. (12) for the energy determined by eq. (6) is found from the relation:

$$S_\mu(E,600) =\ 4.13\times\left(\frac{E}{10^{19}\ \mathrm{eV}}\right)^{0.935}. \tag{14}$$

## 2.3. Examples of typical showers

For the present analysis, we selected EASs with $E$ = (2–12.5)×10$^{18}$ eV and $\theta \leq 60°$, which were registered during the observation period 1978–2018. The final sample includes 1197 and 690 showers in the energy range (2–4)×10$^{18}$ eV and (4–12.5)×10$^{18}$ eV with the axes located inside the central ring of the array with a radius of 1000 and 1300 m, respectively. This allows us to use events with the maximum possible number of triggered MDs in each specific shower. Below, EASs with different muon contents are shown as an example. Fig. 2 illustrates the case when the experimental results are consistent with the hypothesis that this shower was initiated by a primary proton. Fig. 3 and Fig. 4 show events with anomalously high and anomalously low muon contents, respectively (see peaks "D" and "X" in Fig. 6 below).

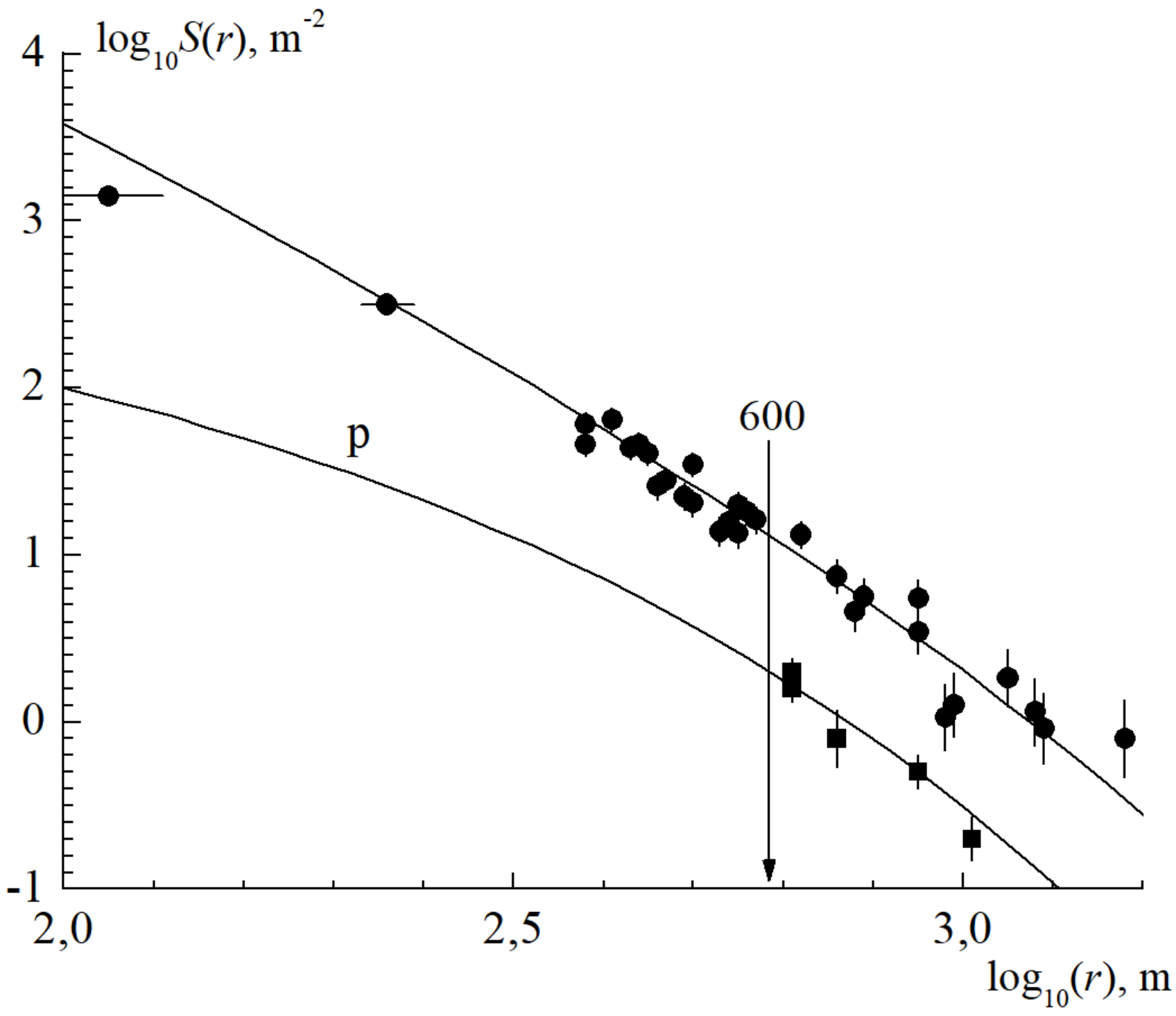

**Fig. 2.** The LDF of shower # 102372 (25.04.1988) with $E$ = 5.37×10$^{18}$ eV, $\theta$ = 9.2°, $S_{600}^{\mathrm{exp}}(E,\theta) = 13.9 \pm 0.3$ and $S_{\mu}^{\mathrm{exp}}(E,\theta,600) = 2.14 \pm 0.36$ m$^{-2}$ (the shower axis was located 470 m from the center of array). Circles and squares are responses of SDs and MDs, respectively. Curves are approximations (9) and (12) (p), calculated according to the model QGSjet-II.04 [11] for the primary proton.

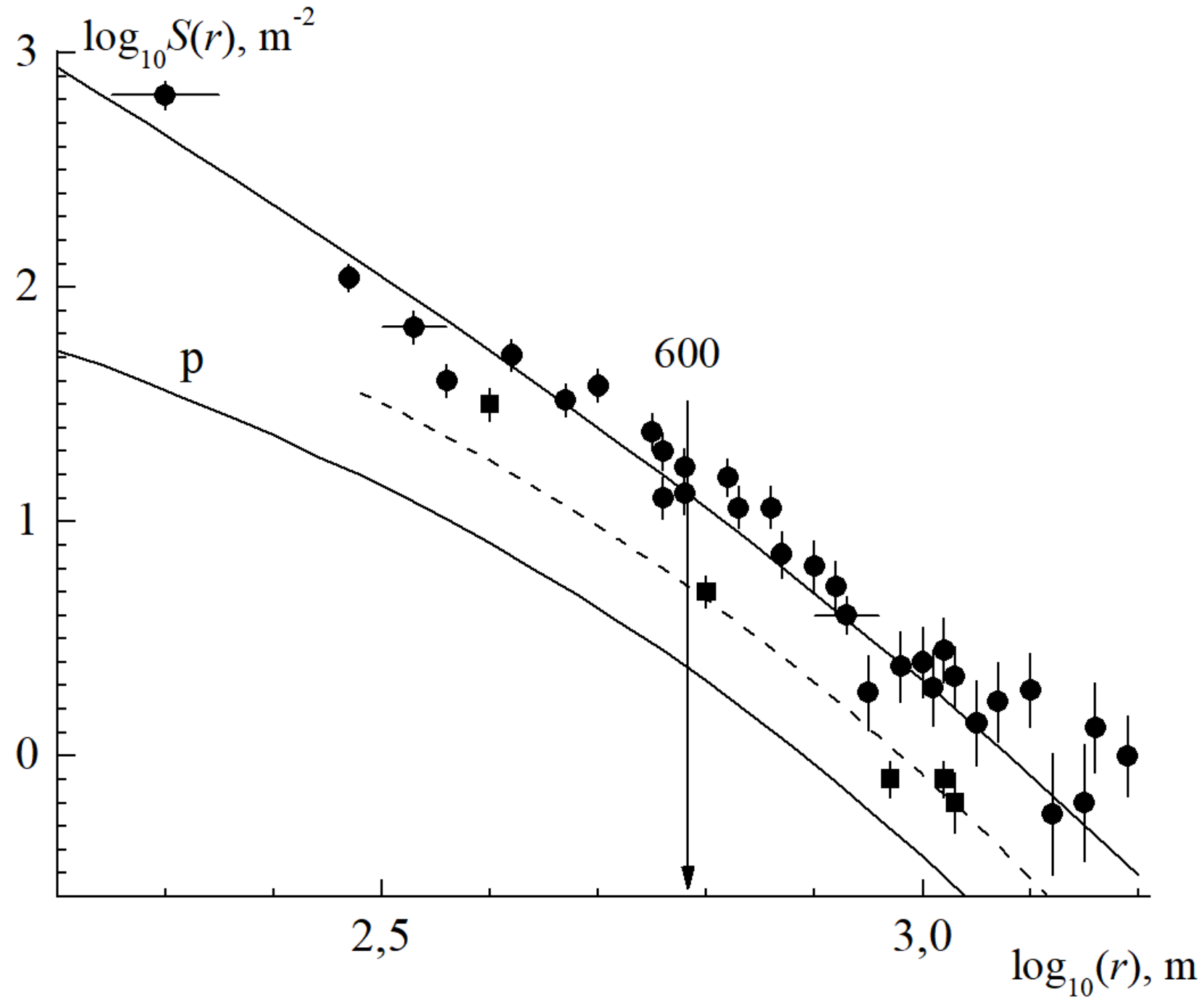

**Fig. 3.** The LDF of shower # 1089094 (25.12.2002) with $E$ = 6.31×10$^{18}$ eV, $\theta$ = 24.2°, $S_{600}^{\exp}(E,\theta)$ = 13.6±1.9 and $S_{\mu}^{\exp}(E,\theta,600)$ = 5.37±0.93 m$^{-2}$ (the shower axis was located 602 m from the center of array). Circles, squares and solid lines mean the same as in Fig. 2. The dashed curve is the approximation (12) multiplied by a factor of 2.24.

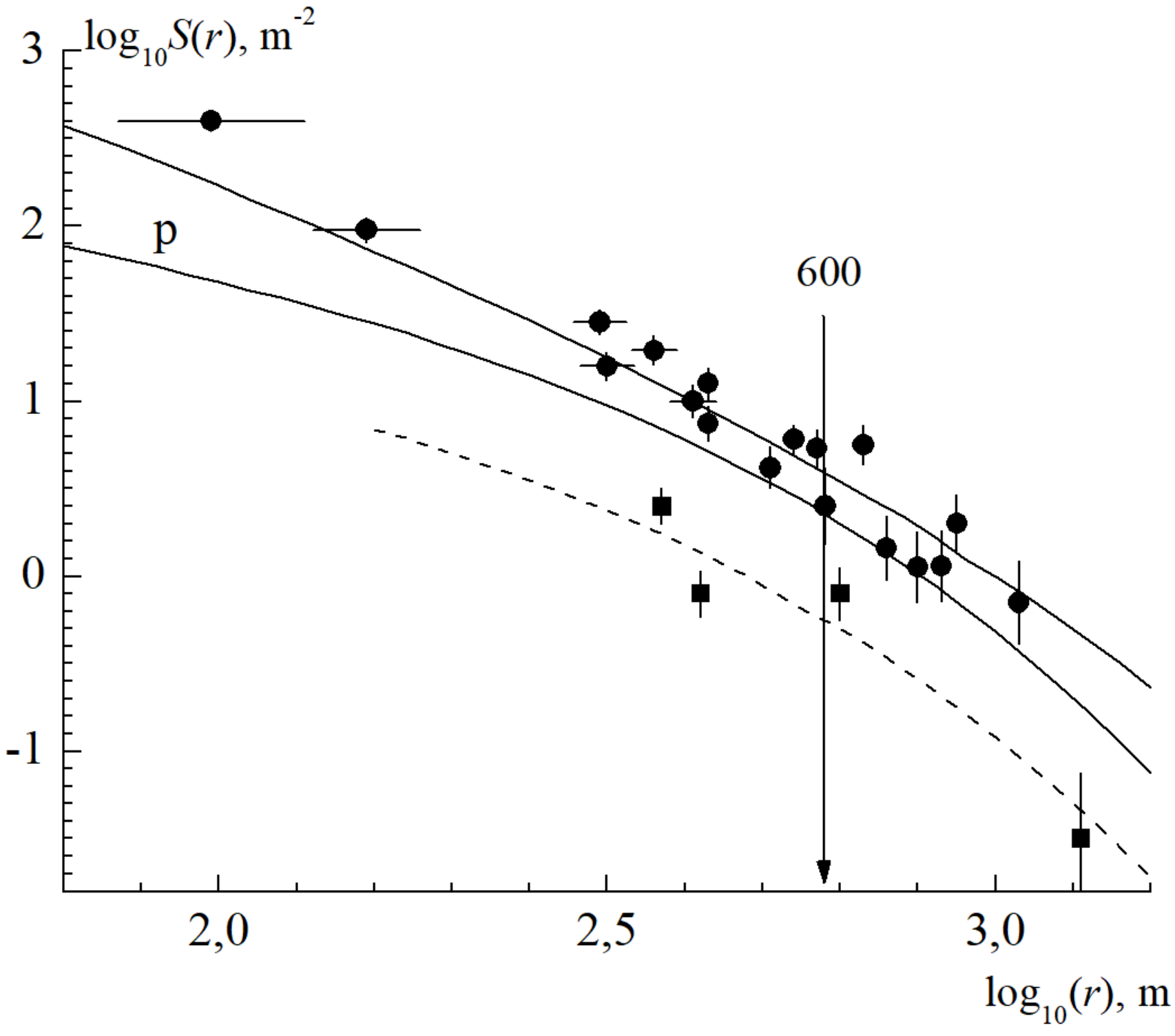

**Fig. 4.** The LDF of shower # 913960 (13.06.2001) with $E$ = 6.02×10$^{18}$ eV, $\theta$ = 53.3°, $S_{600}^{\exp}(E,\theta)$ = 4.0±0.9 and $S_{\mu}^{\exp}(E,\theta,600)$ = 0.58±0.19 m$^{-2}$ (the shower axis was located 406 m from the center of array). Circles, squares and solid lines mean the same as in Fig. 2. The dashed curve is the approximation (12) divided by a factor of 4.

### 2.4. Method of estimation of the CR mass composition

The method for estimating the CR mass composition [31] based on comparison of the measured response density $S_{\mu}^{\mathrm{exp}}(E,\theta,r)$ in a shower with energy $E$ and zenith angle $\theta$ at axis distance $r$ with the expected value $S_{\mu}^{\mathrm{sim}}(p,E^{*},\theta,r)$ calculated using the QGSjet-II.04 [11] for primary proton with a given energy $E^{*}$ and the same $\theta$ and $r$. The calculations showed that the expected muon density in showers with $\log_{10}E \geq 18.3$ eV and $\theta \leq 60°$ can be found from the relation:

$$S_{\mu}^{\mathrm{sim}}(p,E,\theta,r) = S_{\mu}^{\mathrm{sim}}(p,E^{*},\theta,r)\,\frac{S_{600}^{\mathrm{exp}}(E,\theta)}{S_{600}^{\mathrm{sim}}(p,E^{*},\theta)}, \tag{15}$$

where $E^{*} = 10^{19}$ эВ. The value $S_{600}^{\mathrm{sim}}(p,E^{*},\theta)$ characterizes the zenith-angular dependence of the density of the SD responses at $r$ = 600 m. In a vertical shower it is connected via eq. (6) to the energy $E^{*}$ assumed in the calculation, which, within about 6%, does not contradict the estimation of the primary energy obtained at the Yakutsk array [34].

Next, we will consider the correlations of the densities $S_{\mu}^{\mathrm{sim}}(p,E,\theta,r)$ at $r$ = 600 m with the values measured in the experiment. At this distance, the densities of SD and MD have the maximum statistics of showers and are found with the best accuracy. The algorithm for the sequence of calculations of the value (15) is demonstrated in Fig. 5 for the events shown above in Figs. 2–4. The digit ”1” corresponds to Fig. 2. The dark diamond shows the density of $\log_{10}S_{600}^{\exp}(5.37{\times}10^{18}, 9.2°) = 1.14$, which is less than the calculated value of $\log_{10}S_{600}^{\rm sim}(p, 10^{19}, 9.2°) = 1.40$ by Δlog10[$S_{600}$(9.2°)]=1.14–1.41= –0.27. In this case, the value of (15) is equal to:

$$\log_{10}S_\mu^{\rm sim}(p, 5.37{\times}10^{18}, 9.2°, 600) = \log_{10}S_\mu^{\rm sim}(p, 10^{19}, 9.2°, 600) + \\ +\Delta\log_{10}[S_{600}(9.2°)] \ = \ 0.61 - 0.27 \ = \ 0.34, \quad (16)$$

which, within the experimental errors, agrees with the measured muon response density:

$$\log_{10}S_\mu^{\exp}(5.37{\times}10^{18}, 9.2°, 600) = 0.33 \pm 0.07. \quad (17)$$

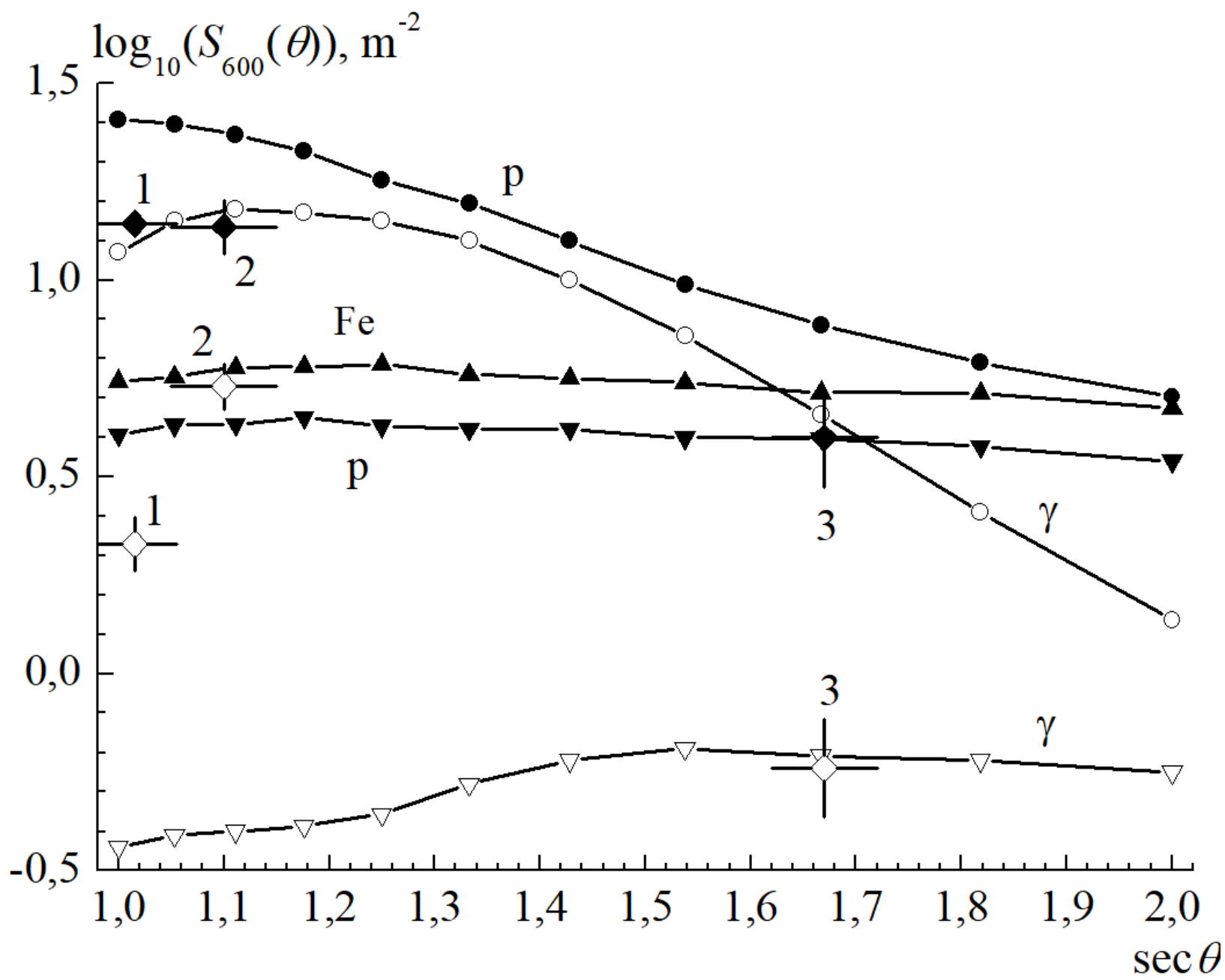

**Fig. 5.** Zenith-angular dependence of the SD and MD (threshold energy $E_{\mu}$ = 1.0×sec$\theta$ GeV) responses at $r$ = 600 m for showers with the energy $E^{*} = 10^{19}$ eV, calculated according to the model QGSjet-II.04 [11] for the primary proton (*p*), iron nuclei (Fe) and gamma photon ($\gamma$). The solid diamonds are SD data, open diamonds are MD data.

Densities (16) and (17) correspond to the peak “p” in Fig. 6.

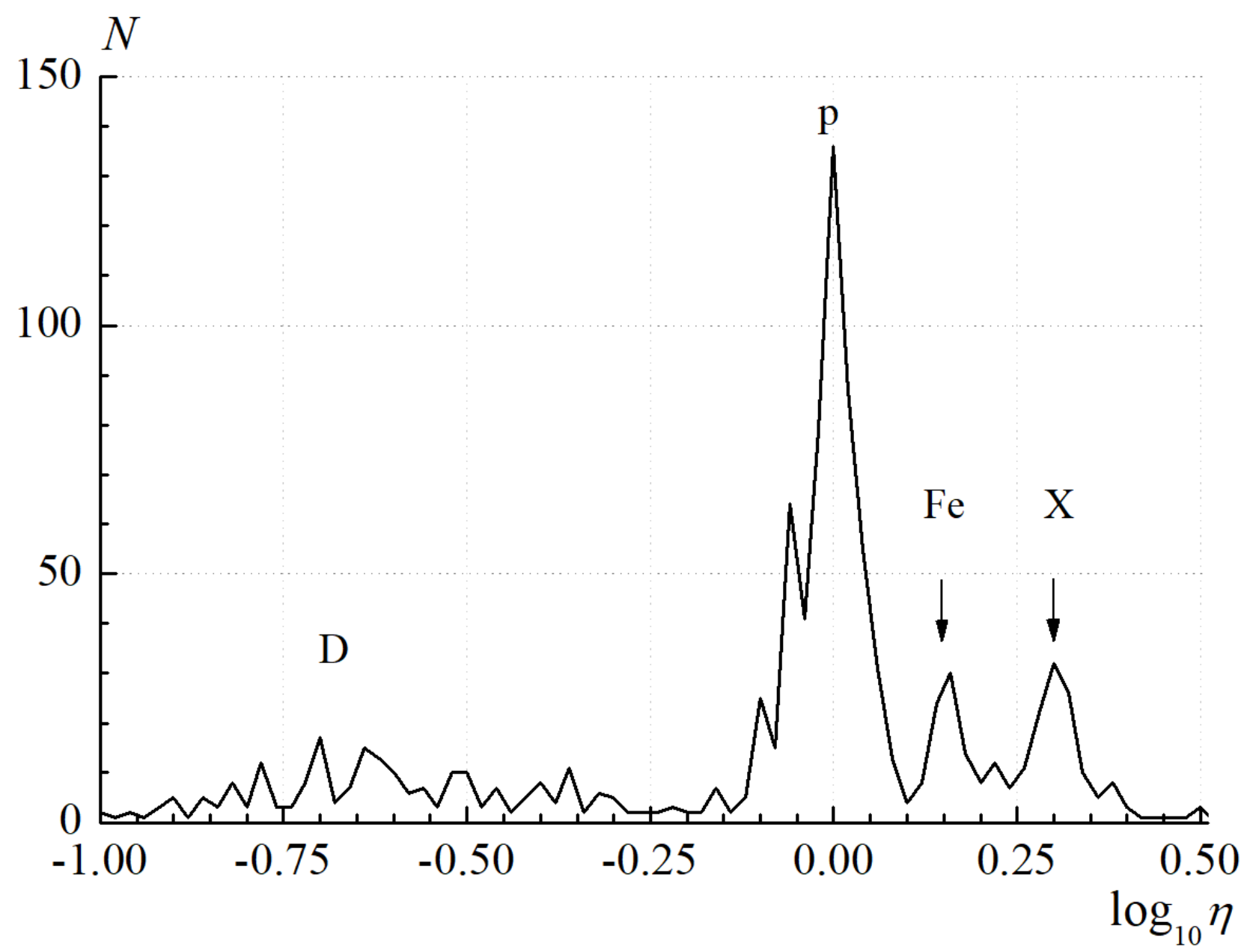

**Fig. 6.** Distribution of ratios of the muon response densities $\eta$ (2) at a distance from the axis $r$ = 600 m, measured experimentally $S_\mu^{\exp}(E, \theta, r)$ and calculated $S_\mu^{\rm sim}(p, E, \theta\ r)$ according to the QGSjet-II.04 model [11] for primary protons in showers with energy $E$ and zenith angle $\theta$.

The next shower on Fig. 3 contains an anomalously high number of muons and is related to the peak “*X*” in Fig. 6. Its measured density $\log_{10}S_{600}^{\exp}(6.31{\times}10^{18}, 24.2°) = 1.13$ is less than the calculated one $\log_{10}S_{600}^{\rm sim}(p, 10^{19}, 24.2°) = 1.37$ by Δlog10[$S_{600}$(24.2º)]=1.13–1.37=–0.24. In this case, the expected value (15) is:

$$\log_{10}S_\mu^{\rm sim}(p, 6.31{\times}10^{18}, 24.2°, 600) = \log_{10}S_\mu^{\rm sim}(p, 10^{19}, 24.2°, 600) + \\ +\Delta\log_{10}[S_{600}(24.2º)] = 0.61\text{–}0.24 = 0.37. \quad (18)$$

It is less than the density $\log_{10}S_\mu^{\exp}(6.31{\times}10^{18}, 24.2°, 600) = 0.73{\pm}0.07$ measured in the experiment by the value:

$$\Delta\log_{10}[S_\mu\ (6.31{\times}10^{18}, 24.2°, 600)] = 0.73\text{–}0.37 = 0.36{\pm}0.07. \quad (19)$$

## 3. RESULTS AND DISCUSSION

### 3.1. CR mass composition

Fig. 6 shows the correlations of pairs of muon densities (2) for a sample of 1887 EASs with energies $E = (2–12.5)\times10^{18}$ eV and angles $\theta \le 60°$. The value "0" on the X-axis corresponds to the location of the proton peak for the QGSjet-II.04 model [11], relative to which the Fe and X peaks have shifts:

$$\langle \Delta_{p-\mathrm{Fe}} \rangle = \log_{10}\left[\frac{S_{\mu}^{\mathrm{sim}}(Fe, E, \theta, r)}{S_{\mu}^{\mathrm{sim}}(p, E, \theta, r)}\right] \approx 0.15\,, \tag{20}$$

$$\langle \Delta_{p-\mathrm{X}} \rangle = 2 \times \langle \Delta_{p-\mathrm{Fe}} \rangle \approx 0.30. \tag{21}$$

The value (20) corresponds to the expected correlation of pairs (2) in EAS from iron nuclei. It is derived by averaging the difference in muon densities in Fig. 5 over the entire range of zenith angles. The second shift (21) refers to showers with an anomalously high muon content. And, finally, the broad peak "D" is caused by showers with a very low muon content. These peaks are located with the 68% confidence level at average values:

$$\langle \eta(p) \rangle = 1.00^{+0.08}_{-0.06}, \tag{22}$$

$$\langle \eta(\mathrm{Fe}) \rangle = 1.41^{+0.05}_{-0.05}, \tag{23}$$

$$\langle \eta(\mathrm{X}) \rangle = 2.26^{+0.06}_{-0.07}, \tag{24}$$

$$\langle \eta(\mathrm{D}) \rangle = 0.2^{+0.13}_{-0.06}, \tag{25}$$

which can be directly related to the different composition of primary particles. Table 1 presents differentiated data on the probable CR composition. It is evident that the proportion of protons in this sample (45 ± 2) % and changes little. The proportion of iron nuclei is (10 ± 3) %. The contribution of X-events remains almost unchanged at about (12 ± 3) %. Finally, D-events make the second largest contribution at (33 ± 2) % to the total CR flux.

Let's find the average value:

$$\langle \eta \rangle = w_p \langle \eta(p) \rangle + w_{\mathrm{Fe}} \langle \eta(\mathrm{Fe}) \rangle + w_{\mathrm{X}} \langle \eta(\mathrm{X}) \rangle + w_{\mathrm{D}} \langle \eta(\mathrm{D}) \rangle = 0.81 \pm 0.12, \tag{26}$$

from which one could erroneously conclude that CRs consist mainly of protons alone. We have come to such conclusion many times before (see, for example, [15–17, 34–38]). From the Table 1 one can see that the estimation of primary particles composition using any average characteristics of EAS could be wrong due to the significant fraction of muon-deficient and muon-excessive events presented in shower samples, which can significantly distort the observed picture.

**Table 1:** The number of EASs in the energy intervals with a step $\Delta \log_{10}E = 0.1$ (in the second column are average energies in these intervals).

| Bin | $\log_{10}E$ | $p$ | $w_p$ | Fe | $w_{Fe}$ | X | $w_X$ | D | $w_D$ | Total |
|---|---|---|---|---|---|---|---|---|---|---|
| 1 | 18.35 | 215 | 0.37 | 65 | 0.11 | 62 | 0.10 | 244 | 0.42 | 586 |
| 2 | 18.45 | 147 | 0.47 | 36 | 0.12 | 20 | 0.06 | 107 | 0.34 | 310 |
| 3 | 18.55 | 145 | 0.48 | 23 | 0.08 | 26 | 0.09 | 107 | 0.35 | 301 |
| 4 | 18.65 | 104 | 0.59 | 13 | 0.07 | 26 | 0.15 | 34 | 0.19 | 177 |
| 5 | 18.75 | 75 | 0.56 | 5 | 0.04 | 23 | 0.17 | 35 | 0.27 | 138 |
| 6 | 18.85 | 94 | 0.45 | 17 | 0.08 | 34 | 0.17 | 63 | 0.30 | 208 |
| 7 | 18.95 | 41 | 0.40 | 19 | 0.18 | 21 | 0.21 | 22 | 0.21 | 103 |
| 8 | 19.05 | 40 | 0.62 | 8 | 0.12 | 11 | 0.18 | 5 | 0.08 | 64 |
| | Total | 861 | 0.45 | 186 | 0.10 | 223 | 0.12 | 618 | 0.33 | 1887 |

Figure 7 shows the weights of the probable composition of CRs with energies $10^{18.3 - 18.6}$ eV (Table 1), obtained from the annual data during 1978 - 2016. These samples provide a number of interesting details. It can be seen that the components "p" and "D" in most cases have a similar share in the total flux of primary particles at the level of ≈40%. The contributions of the “Fe” and "X" components are also similar, at ≈ (5–10) %, possibly due to a shared generation mechanism. Note, the synchronous increase in the contributions of the "Fe" and "X" pair during the 1999–2002, which was accompanied by a decrease in the weights of the other pair, "p" and "D." By 2002, the fractions of all four cosmic ray components in Fig. 6 had roughly equalized. Subsequently, over the next 2–3 years, this dynamic picture returned to its original state. So far, this remains difficult to understand. One can only speculate that, from time to time, some grand explosive processes occur in the depths of the Universe. The nature of the primary "D" and "X" particles also remains far from being understood.

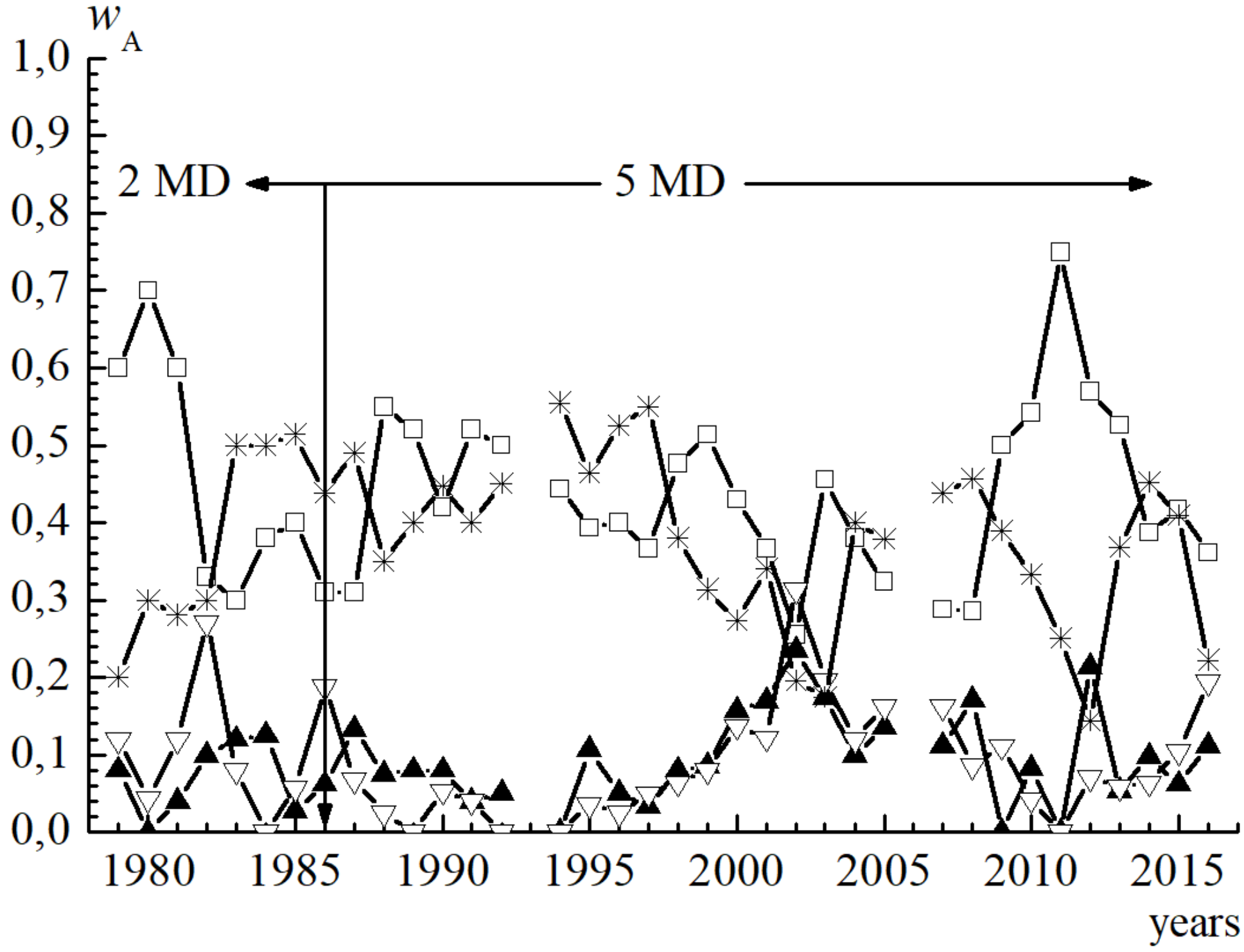

**Fig. 7.** Weight values of the CRs composition (Table 1) in the energy range $E = 10^{18.3 - 18.6}$ эВ according to observations by the Yakutsk EAS array made in 1978 – 2016. □ – protons; ∗ – “D”- component; ∇ – iron nuclei; ▲ – “X”- component. Arrows denotes the periods when 2 and 5 MD was in operation.

### 3.2. D-component of CRs

The case shown in Fig. 4 is consistent with the hypothesis that this component (if not entirely, then at least partially) could be considered a candidate for the role of primary ultra-high-energy gamma rays. This event in Fig. 5 marked by digit 3. One can see that the measured parameters $S_{600}^{exp}(E, 53.3°)= 4.0 \pm 0.9$ and $S_{\mu}^{exp}(E, 53.3°, 600) = 0.58 \pm 0.19$ м$^{-2}$ are close to the expected values $S_{600}^{sim}(\gamma, E, 53.3°)$ and $S_{\mu}^{sim}(\gamma, E, 53.3°, 600)$ calculated for the primary gamma ray with energy $\log_{10}(E^{*}) = 19.0 - 0.05 = 18.95$ eV. The energy $E = 6.02\times10^{18}$ эВ (this event is included in bin 5 in the Table 1) found using measured $S_{600}^{exp}(E, 53.3°)$ by equations (6)–(8) is lower than $E^{*} = 8.91\times10^{18}$ eV by factor 1.48. On the Fig. 5 one could see that for EASs with the different zenith angles the underestimation the energy of the primary gamma photon will be different. The smallest and largest underestimations occur at $\sec\theta \approx 1.25$ and 2.0, respectively, and are $10^{0.085} = 1.22$ and $10^{0.57} = 3.72$ times.

### 3.3. Estimation of the accuracy of muon measurements

Fig. 8 shows the correlation parameters (2) obtained using the QGSjet-II.04 model [11]. They include samples for 80 random vertical showers with an energy of $10^{18.5}$ eV. Calculations have shown that these distributions are almost independent of the shower zenith angle. This is due to the very nature of the development of EAS, in which the muon densities $S_{\mu}(p, E, \theta, 600)$ in inclined events change weakly (Fig. 5). It can be seen that the peaks in Fig. 6 are comparable in width with the peaks in Fig. 8. This suggests that the experimental data obtained above do not distort the real physical picture of the mass composition of CRs and deserve a certain level of trust.

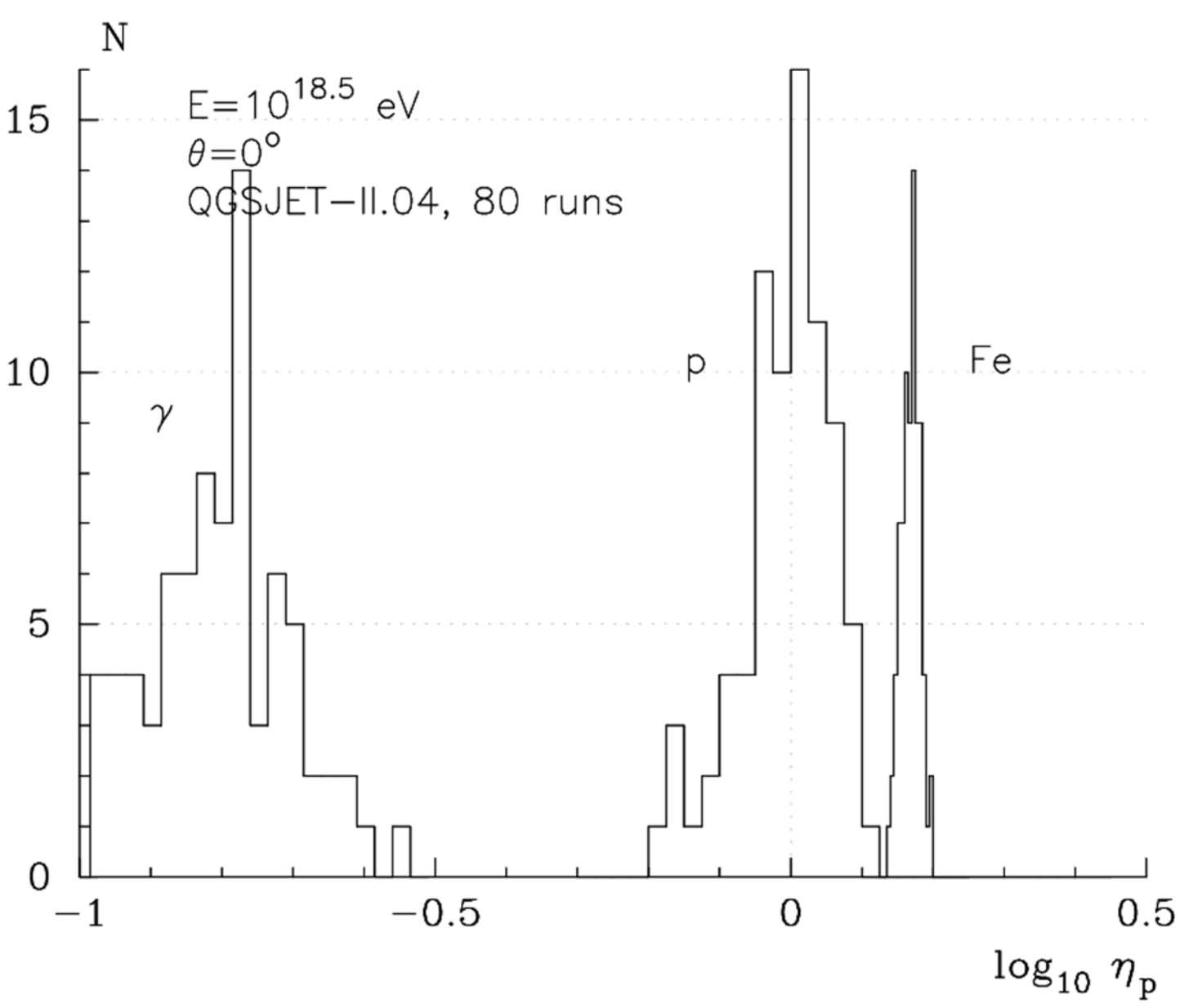

**Fig. 8.** Results of modeling muon densities.

## 4. CONCLUSION

The results of the CR mass composition estimation using the muon correlation method [31] for events with energies $E$ = (2–12.5)×$10^{18}$ eV are presented above. Comparison of the MD responses was performed between experimental data and simulation results obtained using the QGSjet-II.04 model [11] for given energy $E^{*}$ = $10^{19}$ eV. Correlation (2), where the experimentally obtained muon densities at the axis distance $r$ = 600 m in EAS zenith angle $\theta \leq 60°$ were compared with the corresponding calculated values, served as an identifier of the CR mass composition. Figure 6 shows the distribution of correlations (2). It includes 1887 events with the high quality of muon data. This distribution has several clearly defined peaks. The first (22) is undoubtedly formed by primary protons. The second peak (23) can be attributed to iron nuclei. The third (24) contains events with anomalous excess of muon content. Its origin is yet to be unraveled. The fourth peak (25) includes a significant fraction of EAS with anomalously low muon content. There are some indications (see Fig. 4 and Fig. 7) that it may be due to the presence of a considerable fraction of gamma rays in the primary flux of ultra-high energy CRs. In our opinion, the results obtained above are potentially significant for a deeper understanding of the CR composition in the specified region of primary energy. We will continue studying the CR composition using the muon correlation method [31] at lower primary energies, where the Yakutsk EAS array has significantly better statistics.

The authors express their gratitude to the staff of the Yakutsk EAS array. This work was carried out within the framework of a state assignment (registration number #122011800084-7) using the data obtained at The Unique Scientific Facility “The D. D. Krasilnikov Yakutsk Complex EAS Array” (https://ckp-rf.ru/catalog/usu/73611/).